# The Evolution of Alq$_3$ Films Exposure to Air

Min Wang, Yan Zhou, Sui Kong Hark, Xi Zhu

School of Science and Engineering, The Chinese University of Hong Kong, Shenzhen, Shenzhen, Guangdong, 518172

**Abstract:**

Small molecular solar cell becomes more stable when a thin tris-(8-hydroxyquinoline) aluminum (Alq$_3$) buffer layer instead of bathocuproine (BCP) is inserted between the active layer and electrode. In this work, we introduce a single layer device (ITO / Alq$_3$ / Au) exposure to air to investigate the role of Alq$_3$ in organic solar cells. The large PL intensity and undetectable Raman peaks of 8-hydroxyquinoline (8-Hq), a degradation product of Alq$_3$ through chemical reaction, indicate that the degradation of Alq$_3$ is not severe even after a few hundred hours exposure. Thus the Alq$_3$ buffer layer gives a good protection of the active layers against oxygen\water permeation comparing with BCP. However, the small quantity of degraded Alq$_3$ products can enhance the current density (after 15 h) in single layer Alq$_3$ device, which is reflected in the decrease of parallel resistance $R_p$. Therefore, due to the enhanced charge transport and the long term better protection from ambient oxygen/water, the lifetime of small molecular solar cell with Alq$_3$ buffer layer is improved.

*Corresponding author
Email:ustc0200@gmial.com

## 1. Introduction:

Organic photovoltaic devices have gained a considerable interest in the last few years due to their potential advantages of light weight, flexibility, and low-cost [1] . From the first breakthrough by Tang [2] in 1986, different materials and structures have been designed to improve the power conversion efficiency ($\eta$) and the stability of organic solar cells. The reported efficiency has increased considerably from 1% [2] to more than 5.5% [3] for small molecules heterojunction devices recently. The improvements was realized by Peumans and Forrest [4] through the introduction of a buffer layer-a bathocuproine (BCP) thin film, which was placed between $C_{60}$ and cathode in copper phthalocyanine (CuPc) / fullerene ($C_{60}$) cells. However, BCP-based cells are instable due to the rapid crystallization of BCP [4], especially in the presence of moisture. Then a new buffer material tris-(8-hydroxyquinoline) aluminum ($Alq_3$) is introduced to improve the stability [5,6]. The improvements by $Alq_3$ are attributed to the more effective blocking of cathode diffusion than BCP, and a stronger oxygen blocking effect to protect $C_{60}$, comparing with that of BCP. However, the effects of the degradation of $Alq_3$ itself have not been considered in organic solar cells, though $Alq_3$ and its degradation have been widely investigated in organic light emitting diodes (OLEDs) [7-18]. Since $Alq_3$ acts as a buffer layer for solar cells, it is important to understand the charge transport mechanism and the recombination process in $Alq_3$ films. Therefore we design single $Alq_3$ layer devices for our discussion. Gold cathode instead of Al is used to reduce the $Al_2O_3$ effect. Electronic impedance, which is widely used to study carrier injection, charge transport and recombination processes in OLEDs [19-23], is adopted here for studying single-layer $Alq_3$ devices. In this work, we focus on the evolution of device exposure in air to investigate its effects on the efficiency and lifetime of $Alq_3$-based solar cells. The device is characterized by frequency-dependent impedance spectroscopy, current-voltage characteristics, Raman and the evolution of photoluminescence as well.

## 2. Experiment

$Alq_3$ was purified by sublimation and the film was deposited on indium tin oxide (ITO) glass substrate. Then gold was vapor deposited as the cathode. The devices with the structure of ITO / $Alq_3$ (60 nm) / Au (40 nm) (as shown in the inset of Fig. 3) were fabricated in vacuum under the pressure ~ $10^{-6}$ Pa. The current density-voltage (J-V) curves were measured by Keithley 2400. And the impedances of the devices were studied by impedance gain/phase analyzer (SI1260, Solatron) and dielectric interface (SI1294, Solatron) in the frequency range from 0.1 Hz to $10^6$ Hz. Photoluminescence spectra were recorded using an Aminco Bowman II luminescent spectrometer (Thermo Electron, U.S.A) with an excitation wavelength of 360 nm. Raman RXN1 Analyzer from Kaiser Optical Systems, Inc. (KOSI) with an immersion fiber-coupled MR probe (300 mm length $\times$ 12.5 mm diameter) was used to detect the samples under an excitation of 325 nm.

## 3. Results and discussion

The degradation of Alq$_3$ (60 nm on quartz) happens all the time demonstrated by its photoluminescence (PL) intensity in Fig. 1. The inset shows the normalized PL peak intensity versus time. In the first 15 hours, the intensity decays less than 2%; and after 48 hours, the PL intensity decreases to 94% of its original value; after 300 hours, the PL intensity remains about 65% of the one before exposure. It has been found that no PL can be detected even only 2% of polymer III (chemical structure shown in Fig. 6, which is believed to be the degradation product of Alq$_3$) added into Alq$_3$ [24]. The large PL of our device after a few hundred hours exposure in air means that the degradation of Alq$_3$ is not severe and the protection of active layer against oxidation and hydrolysis is good when Alq$_3$ is applied as a buffer layer between active layer and cathode.

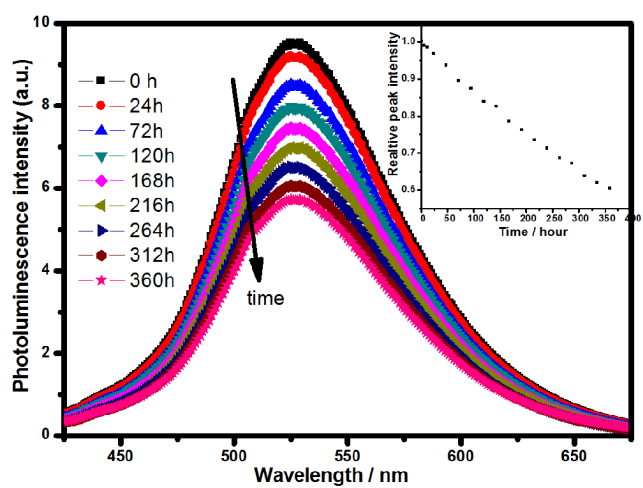

Fig. 1. Photoluminescence intensity of Alq$_3$ film with an excitation wavelength of 360 nm and the inset shows normalized photoluminescence peak intensity v.s. time of exposure to air.

To measure the degradation product of Alq$_3$ (60 nm on quartz), a sample kept in air for more than two weeks is investigated by Raman detection (see Fig. 2). Comparing with the calculated Raman results of Alq$_3$ and 8-Hq based on hybrid density functional B3LYP/6-31G* [25-27] method reported in the Ref. [28], the characteristic Raman peaks of Alq$_3$ located at 504, 526 and 541 cm$^{-1}$ are clearly observed. No peak of 8-Hq (8-hydroxyquinoline and its chemical structure shown in Fig. 6) located at 1278 cm$^{-1}$ (characterized by the $\delta_{COH}$ bending mode) can be detected. Thus the obvious decay of PL and undetectable degradation products mean that these degradation products do exist but their quantity is very small.

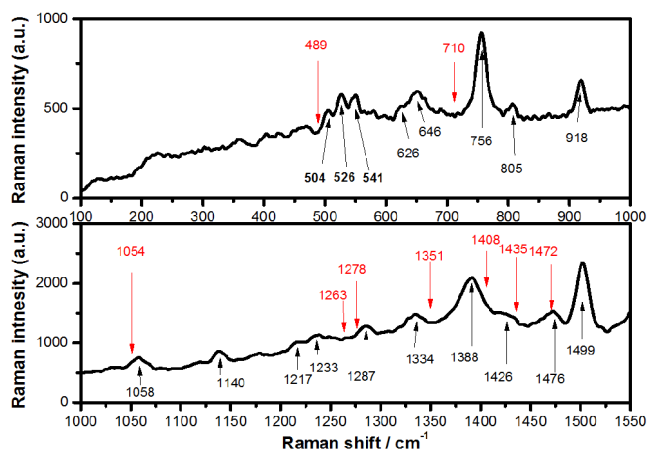

Fig. 2 Energy range (100 ~ 1600 cm$^{-1}$) Raman spectrum of Alq$_3$ film, which kept in air for more than 2 weeks, with an excitation wavelength of 325 nm. The red numbers above the curves depict the characteristic peaks of 8-Hq, and the black ones denote for those of Alq$_3$.

Illuminated by the obvious effects of these small quantities of degradation product of Alq$_3$ on PL, further study of its effect on charge transport, which is more directly on the performance of solar cells, is needed. Then we explore current density-voltage of the device as shown in Fig. 3. The figure shows variations of current density at bias voltage of 1 V versus different exposure time. The current density-voltage curves in the voltage range from -1 V to 1 V (at the exposure times of 0 h, 15 h, 40 h, 50 h, and 135 h) are shown in the insert of Fig. 3. It can be observed that the current density under 1 V decreases at the first 15 hours, then increases quickly till 50 hours, and increases slowly, finally saturates after 135 hours. The largest current density at 1V is about 2.6 mA/cm$^2$, about 20 times larger than the lowest one (0.125 mA/cm$^2$). Hence, the device has a large enhancement of charge transport after it has been exposed to air for certain hours (longer than 15 hours). Obviously, the increase of the current can be attributed to either the contact resistance or the change of Alq$_3$ film itself due to degradation. In order to find out which factor is more important, we characterize the device by impedance spectroscopy during its degradation.

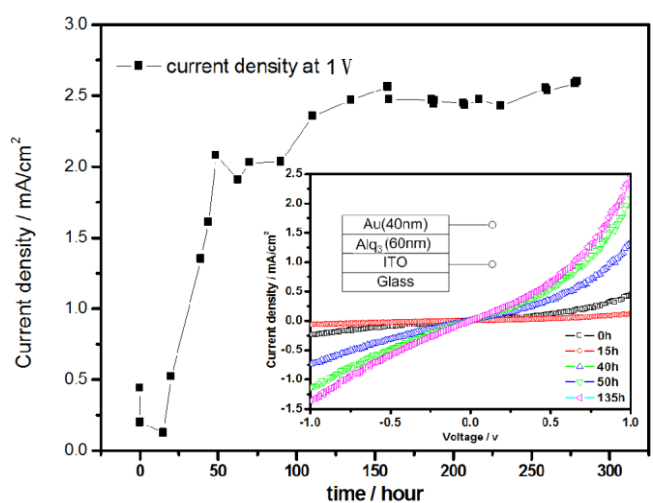

Fig. 3 Variations of current density at bias voltage of 1 V versus different exposure time. J-V characteristics of Alq$_3$ film exposure in air for different hours (0 h, 15 h, 40

h, 50 h, and 135 h) and schematic drawing for the fabricated sandwich device ITO / Alq$_3$ (60 nm) / Au (40 nm) are shown in the inset.

The complex impedance of our device can be expressed in terms of the real and imaginary components $Z'$ and $Z''$:

$$Z = Z' + jZ'' = |Z|^{j\theta}, \tag{1}$$

where $|Z|$ and $\theta$ are the magnitude and the phase of the impedance, respectively.

Fig. 4 shows the evolutions of (a) magnitude $|Z|$ and (b) phase $\theta$ of the device as functions of the frequency from 0.1 Hz to $10^6$ Hz at different exposure time (0 h, 15 h, 40 h, 50 h, and 135 h). As Fig. 4 (a) shows, the magnitude of the impedance $|Z|$ remains constant at the low frequency range, and then decreases at the high frequency range. The corresponding phase $\theta$ is around zero in the low frequency region, decreases to about -90$^o$ in the middle frequency region, and then gradually increases in the high frequency region. At different exposure times, the shapes of $|Z|$ are similar, except the magnitudes at low frequencies. Herein we find the magnitude of $|Z|$ is time dependent, which increases in the first 15 hours and then decreases before converging after 50 hours. The corresponding phase $\theta$ has a similar trend as well. We employ an equivalent circuit model of our device shown in the inset of Fig. 4 (a) to fit the impedance data. The circuit consists of a series resistance R$_s$, a parallel resistance R$_p$ and a parallel capacitance C$_p$. Then, the complex impedance Z can be written as

$$Z = Z' + jZ'' = R_s + \frac{R_p}{1 + j\omega R_p C_p} = R_s + \frac{R_p}{1 + j\omega\tau}, \tag{2}$$

where $\tau$ is a characteristic time constant given by $R_p C_p$.

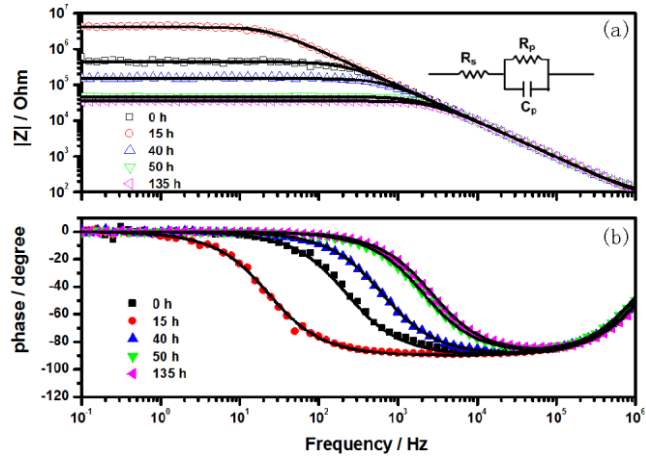

Fig. 4 Changes in impedance spectra for 60-nm-thick film of Alq$_3$ measured and fitted in different exposure times. Symbols are measured data and the solid lines are the fitting data based on equivalent circuit shown in the inset. (a) open symbols, |Z|; (b) solid symbols, phase φ.

The solid lines in Fig. 4 (a) and (b) are based on Eq. (1) and (2) fit the experimental data well, which imply that this equivalent circuit is appropriate for the device structure. The parameters for the equivalent circuit of the device are shown as a function of exposure time in Fig. 5 (a) $R_s$ and $C_p$ and (b) $R_p$. The series resistance $R_s$ might come from the ITO electrode, the gold electrode and the interface resistance between the electrode and the organic layer. Since $R_s$ remains constant around 90 Ω, nearly the same as the resistance of pure ITO, gold electrode and interface resistance are not the reasons $R_s$ originates from. The capacitance $C_p$ is around $10^{-9}$ F and changes a little with different exposure times, indicating little influences of the degradation of Alq$_3$ on the capacity of the device. Unlike the small changes of $R_s$ and

$C_p$, $R_p$ varies a lot during the whole degradation process. $R_p$ increases dramatically to the highest value at the first stage (first 15 h), and decreases quickly in the following 35 hours, then remains constant after 50 hours. The largest $R_p$ is about $4\times10^6\,\Omega$, which is one-order larger than original one (around $4.3\times10^5\,\Omega$) and two-order larger than the lowest parallel resistor ($\sim 3\times10^4\,\Omega$). The change of current density in Fig. 3 thus can be ascribed to the variations of the parallel resistor $R_p$, indicating an improved charge transport. Then the collection of electron by cathode through $Alq_3$ would be improved in organic solar cells. This process would slow down the decrease of efficiency and lengthen the lifetime finally. We find that the reduced current and the corresponding $R_p$ in the first 15 hours take negative effect for the electron transport. However, because of the existence of $Alq_3$, the active layer is well protected and the "negative effect" mentioned above is weaken, thus a number of generated electrons can transport to the cathode freely.

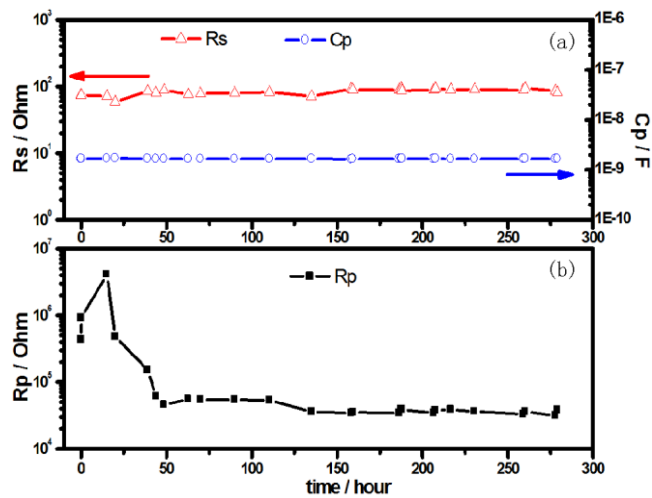

Fig. 5 Variations of fitting parameters ($R_s$, $C_p$ and $R_p$) of the equivalent circuit versus time. (a) open triangle for $R_s$, and open circle for $C_p$; and (b) solid square for $R_p$.

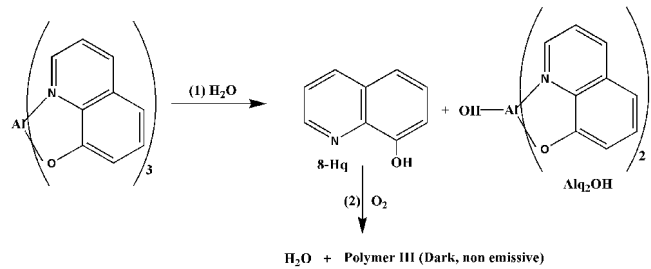

Fig. 6 The scheme for the process of the degradation of Alq$_3$.

In order to explain the above mentioned experimental results, the degradation of Alq$_3$ due to chemical reaction with two crucial processes [24,29,30] are taken into account, including (1) hydrolysis of Alq$_3$ according to a ligand-exchange mechanism to replace the quinolates ligands with a hydroxyl group (Alq$_2$OH), releasing a free 8-hydroxyquinoline ligand (8-Hq); and (2) the oxidative of 8-Hq to form the non-emissive polymer III (the scheme shown in Fig. 6). These three important reaction products: 8-Hq, Alq$_2$OH and polymer III may play different roles to the effects of charge transport. The band gaps of Alq$_3$, 8-Hq and polymer III measured by absorption are 2.67 eV, 3.39 eV and 1.53 eV, respectively [24]. Comparing with Alq$_3$, blue shift of the band edge and peak position for Alq$_2$OH obtained by B3LYP/6-31G* method in the Ref. [31]. The band gap of Alq$_2$OH can be estimated as 3.18 eV while the calculated absorption spectra for Alq$_3$ is 2.95 eV [31]. Because of the large band gaps of 8-Hq and Alq$_2$OH, when mixed with Alq$_3$, the electron transport processing is hindered spontaneously. While due to the smaller band gap of polymer III, the transportation of the electrons is facilitated. Therefore, herein based on the observation of our experiments, we concluded that in the early stage, the 8-Hq and Alq$_2$OH are dominated and $R_p$ of the device is enlarged as their large band gaps. The effect of polymer III becomes more important then and compensates much of the effects of 8-Hq and $Alq_2OH$, reducing $R_p$. Those two competing effects between 8-Hq / $Alq_2OH$ and polymer III finally converge $R_p$ to a constant.

## 4. Conclusion

The evolution of single layer device (ITO / $Alq_3$ / Au) exposure to air has been investigated to understand the role of $Alq_3$ buffer layer in organic solar cells. Even after a few hundred hours exposure, the large PL intensity and undetectable Raman peaks of 8-Hq indicate the degradation of $Alq_3$ is not severe and the buffer layer gives a good protection of the active layers against oxygen\water permeation comparing with BCP. The small quantity of degradation products of $Alq_3$ can enhance the current density (after 15 h) in single layer $Alq_3$ device, which is reflected in the decrease of parallel resistance $R_p$. Therefore, due to the enhanced charge transport and the long term better protection from ambient oxygen/water, the lifetime of small molecular solar cell with $Alq_3$ buffer layer is improved.

**Acknowledgement:**

This work is financially supported by xxx.

**Figure captions:**

Fig. 1 Photoluminescence intensity of $Alq_3$ film with an excitation wavelength of 360 nm and the inset shows normalized photoluminescence peak intensity v.s. time of exposure to air.

Fig. 2 Energy range (100 ~ 1600 $cm^{-1}$) Raman spectrum of $Alq_3$ film, which kept in air for more than 2 weeks, with an excitation wavelength of 325 nm. The red numbers above the curves depict the characteristic peaks of 8-Hq, and the black ones denote for those of $Alq_3$.

Fig. 3 Variations of current density at bias voltage of 1 V versus different exposure time. J-V characteristics of $Alq_3$ film exposure in air for different hours (0 h, 15 h, 40 h, 50 h, and 135 h) and schematic drawing for the fabricated sandwich device ITO / $Alq_3$ (60 nm) / Au (40 nm) are shown in the inset.

Fig. 4 Changes in impedance spectra for 60-nm-thick film of $Alq_3$ measured and fitted in different exposure times. Symbols are measured data and the solid lines are the fitting data based on equivalent circuit shown in the inset. (a) open symbols, |Z|; (b) solid symbols, phase φ.

Fig. 5 Variations of fitting parameters ($R_s$, $C_p$ and $R_p$) of the equivalent circuit versus time. (a) open triangle for $R_s$, and open circle for $C_p$; and (b) solid square for $R_p$.

Fig. 6 The scheme for the process of the degradation of $Alq_3$.